\documentclass[a4paper,12pt]{article}
 \usepackage{epsfig,float,floatfig,wrapfig}
 \usepackage[hang,small,ruled,bf]{caption}
\newcommand{\be}{\begin{equation}}
\newcommand{\ee}{\end{equation}}
\newcommand{\beq}{\begin{eqnarray}}
\newcommand{\eeq}{\end{eqnarray}}

\begin{document}
{\Large
\begin{center}
{\bf The role of three-body collisions 
in  \mbox {\boldmath $\phi$} meson production processes near threshold}
\end{center} }

\begin{center}
H.W. Barz and B. K\"ampfer\\ 
Forschungszentrum Rossendorf, 
Pf 510119, 01314 Dresden, Germany\\

\vspace*{10mm}
\end{center}
\vspace*{10mm}
{\bf Abstract:}\\
The amplitude of subthreshold $\phi$ meson production
is calculated using dominant 
tree-level 
diagrams for three-body collisions. It is shown 
that the production can overwhelmingly be 
described by two-step processes. 
The effect of the genuine three-body 
contribution (i.e. the contribution which cannot be factorized) 
is discussed. 
The production rate of  $\phi$ mesons is presented
for proton  induced reactions on carbon. 



\section{Introduction}

Currently there is  much interest in subthreshold production
of heavy mesons in relativistic  nuclear collisions. It is
expected that measurements will reveal the change of the 
properties of mesons in dense and hot nuclear matter. 
Such predictions are made from theoretical reasons 
\cite{Kaplan} and, indeed, hints to a dropping 
mass of anti-kaons 
have been found by comparing recent measurements \cite{laue} 
with calculations  based on transport models \cite{Cassing}. 
For  $\phi$ meson production there are only a few measurements 
\cite{herrmann} in the reactions of Ni+Ni at 1.93 A$\cdot$GeV
and Ru+Ru at 1.69  A$\cdot$GeV.
Although only a limited amount of the 
phase space was accessible in these experiments, 
extrapolations to the full phase space point to a surprisingly 
large production cross section. 

First estimates 
using transport models \cite{chung2}  based on two-particle cross sections,
derived in a simple meson exchange model \cite{chung1},  
seem to underestimate the $\phi$ multiplicity.
One should however take in to account the lack of precise knowledge
concerning elementary $\pi$$\Delta$,  $\Delta$N and $\Delta$$\Delta$ 
collisions. 
In addition multiparticle processes could contribute.
Experimentally, such multi-step processes can be investigated 
by accompanying cluster emission 
in meson production \cite{hmueller}.

In this work we study whether three-body
collisions could remarkably contribute to 
the production of $\phi$ mesons in proton-nucleus and 
heavy-ion collisions.
At threshold such a mechanism could be dominant
because less bombarding energy is required if the
projectile nucleon interacts with two target nucleons instead with 
a single one.  For 'free' nucleons the threshold reduces from 
2.6 GeV to 1.8 GeV. 

The consideration of $\phi$ production is interesting 
since also the K$^-$ production proceeds considerably
through an intermediate $\phi$ meson. In the near-threshold 
proton-proton collision roughly half of the K$^-$ 
mesons come from the $\phi$ decay \cite{DISTO}. 
K$^-$ mesons have been measured
by the KaoS collaboration
at the heavy-ion synchrotron SIS of GSI Darmstadt 
in proton-Au collisions for a bombarding energy
of 3.5 GeV \cite{scheinast}, and the ANKE collaboration envisages the 
K$^-$ measurements for collisions of protons on various light targets
at the cooler synchrotron COSY of FZ J\"ulich 
\cite{ANKE}. For a firm understanding of these
reactions a necessary prerequisite is the theoretical understanding
of the role of the $\phi$ channels.

\begin{figure}
\centerline{
\includegraphics[width=16cm,height=10.cm]{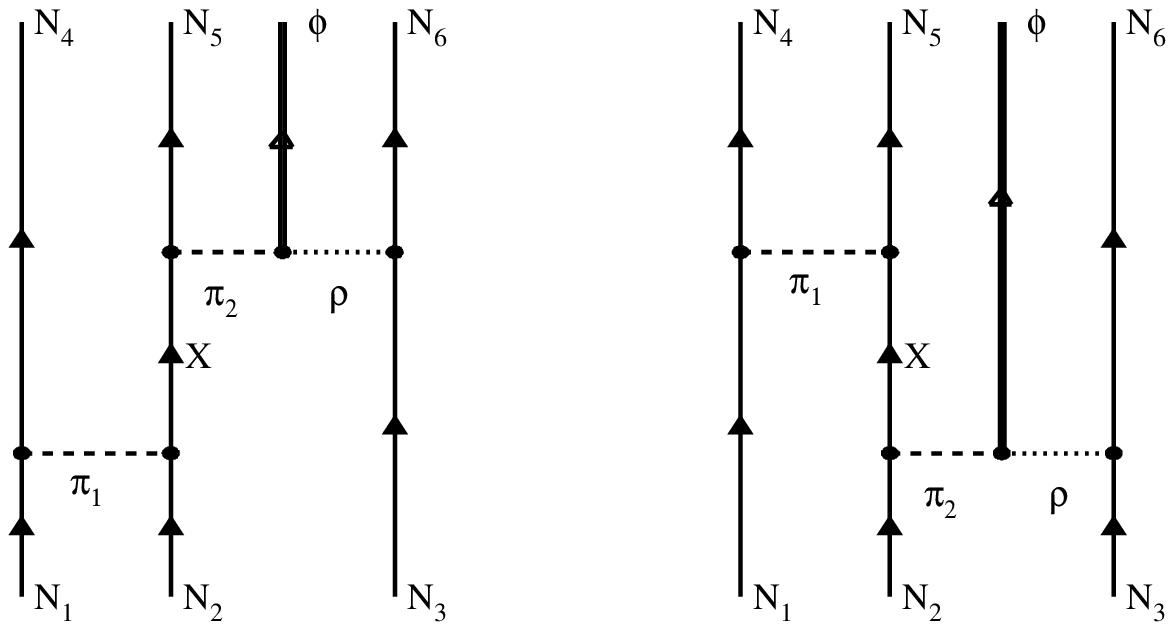}  }
\caption{
Dominating tree-level 
diagrams contributing to the three-body $\phi$ meson
production. Two further diagrams are used in the text, where the 
$\pi_2$ and the $\rho$ lines are interchanged.
\label{graph1} }
\end{figure}

We are going to calculate the cross section 
for the reaction p+2N $\to$ $\phi$ + 3N from the 
invariant amplitude based on tree-level diagrams.
In refs.~\cite{titov1,titov2} several diagrams were explored thoroughly
for the case of the two-nucleon $\phi$ production, i.e. N+N $\to$ 
$\phi$+2N. As shown 
there, the available data can be described by diagrams
with internal meson conversion in a $\pi$$\rho$$\phi$ vertex.
Based on this investigation we restrict ourselves to such types of diagrams
as illustrated in Fig.~\ref{graph1}.
Three nucleons in states $N_1$, $N_2$, $N_3$ act together
via boson exchanges.
Literally in the first diagram, after the interaction of particles 1 and 2,
the intermediate fermion $X$ or pion $\pi_2$ has gathered sufficient
energy to produce the  $\phi$ meson interacting  with the third
nucleon. In the right hand diagram the time ordering is exchanged.

Similar processes are described within the current transport models 
by sequential two-step processes which also allow to accumulate
the energy of several nucleons. In this case, however, the intermediate
particles are real, i.e. they are moving on-shell. 
Recently also effort is made to implement properly the particles' 
off-shell propagation (see ref.~\cite{Leupold}). 
In the diagrams  displayed in Fig.~\ref{graph1}
the intermediate particles $X$ and $\pi_2$ 
can be either on shell or off shell depending on 
the initial and final momenta
of the particles. There are subspaces in the phase space, 
where the intermediate particles move on shell.
It is our aim to investigate these two effects and 
to relate  them to the standard treatment as sequential 
two-step processes.

Many-body collisions have already been treated several times in 
the literature.
In ref.~\cite{danielewicz} this mechanism was applied to analyse
the cumulative backward emission of nucleons. In this region of
the phase-space off-shell contributions become important.
Many-body collisions were also incorporated in the framework 
of transport models \cite{kodoma,batko}.
The results of these investigations show moderate effects; 
here we extend these studies to very subthreshold reactions
for the $\phi$ production.

\section{Elementary cross section}

To calculate the cross section for three-particle collisions 
we will apply the one-boson exchange model
which is often used to parametrize Lorentz invariant cross sections. 
Our approach is strictly based on tree-level diagrams
with effective
parameters adjusted to experimental data.
For our investigations we restrict ourselves to the diagrams 
displayed in Fig.~\ref{graph1}. 
We also consider two further diagram types, where the 
$\rho$ and the $\pi_2$ mesons are interchanged. For the sake of
simplicity in  the present exploratory 
investigation we employ only the  
nucleon propagator for the intermediate particle $X$ and the pion exchange
for the interaction between nucleons. 
Under these propositions the whole set of diagrams we are considering 
consists of  \mbox{4 $\times$  36} individual 
diagrams, where the second factor denotes the number of 
permutations of the respective three fermion lines in the initial and
final states. 

In ref.~\cite{chung2} the interaction Lagrangians for these diagrams 
have been presented which read in standard notation 
\beq
{\cal{L}}_{\pi NN} 
          & =& -\frac{f_{\pi NN}}{m_\pi} \bar{\psi} \gamma^5
              \gamma^\mu \vec{\tau} \psi \partial_\mu \vec{\pi}\;,\\
{\cal{L}}_{\rho NN} 
          & =& -g_{\rho NN} \bar{\psi}
                   \gamma^\mu \vec{\tau} \psi \vec{\rho}_\mu
               - \frac{f_{\rho NN}}{2 m_N} \bar{\psi} \sigma^{\mu\nu}
                 \vec{\tau} \psi \partial_\mu \vec{\rho}_\nu\;,\\
{\cal{L}}_{\pi\rho \phi} 
          & =& \frac{f_{\pi\rho \phi}}{m_\phi} 
                \epsilon_{\mu\nu\alpha\beta} \partial^\mu \phi^\nu
                \partial^\alpha {\vec{\rho}}\,^\beta  \vec{\pi}\;.
   \label{lagran}
\eeq
These equations  contain the nucleon field $\psi$,  
the iso-vector meson fields $\vec{\pi}$, $\vec{\rho}\,^\mu$ and 
the iso-scalar field 
$\phi^\mu$. The Dirac matrices, the
Levi-Civita symbol and the corresponding particle masses 
are denoted by $\gamma^\mu$, 
$\epsilon_{\mu\nu\alpha\beta}$ and $m_i$, respectively.
Greek indices stand for Lorentz indices, and we put $\hbar=c=1$.
Furthermore, a formfactor is needed for each internal line of 
an exchanged meson $i$ coming from or entering 
a vertex $(jkl)$.
We adopt here the standard monopole form factor as a function of the 
square of the transferred four-momentum $q_i^\mu$ and a cut-off
parameter  $\Lambda^i_{jkl}$:
\beq
  F(q_i^2) = \frac{(\Lambda^i_{jkl})^2 - m_i^2}
              {(\Lambda^i_{jkl})^2 - q_i^2}.
\eeq

The following values of the coupling constants and cut-off parameters
have been used: 
$f_{\pi NN}= 0.989$, $g_{\rho NN}=3.718$,
$f_{\rho NN}=6.1g_{\rho NN}$, 
$f_{\pi\rho \phi} = 1.04$;  
     $\Lambda^\pi_{\pi N N}$       \mbox{ = 1.6 GeV, }
     $\Lambda^\rho_{\rho N N}$      \mbox{ = 1.3 GeV, }
     $\Lambda^\pi_{\pi \rho \phi}$  \mbox{ = 0.95 GeV, }
     $\Lambda^\rho_{\pi \rho \phi}$  \mbox{ = 1.2 GeV. }
Most of these values are taken from refs.~\cite{chung2,machleit}
and have been successfully applied in the near-threshold region. 
For  $\phi$ meson production
at an energy of 82  MeV above the production threshold
we obtain a cross section of 0.04 $\mu$b using these parameters. 
This value is smaller than the recently measured 
value \cite{DISTO}. However, 
in ref.~\cite{titov2} it has been demonstrated that 
the agreement between data and calculation
is improved  
by including in addition the $\phi$ meson bremsstrahlung 
from external nucleon legs and the final state interaction.

The cross section for $\phi$ meson production by an incoming
nucleon with momentum $p_1=p_{lab}$  hitting a nucleon with
momentum $p_2$ which is surrounded by another nucleon with 
momentum  $p_3$ in a medium of density
$n_0$ reads
\beq
  \sigma = \frac{1}{4m_N p_{lab}} \, \frac{n_0}{2m_N} \,
           \frac{1}{2} \; \frac{1}{3!}\; \frac{1}{(2\pi)^8}\; 
           \frac{1}{8}\sum_{projections}
           \int d{\bf L}_{inv}
           \mid \sum_{diagrams} {\cal{T}}\mid ^2.  \label{sigma}
\eeq
The first quotient contains the incoming flux with the 
laboratory momentum $p_{lab}$. 
The second quotient is related to the density $n_0$
of the second collision partner; the fact that we     
calculate the cross section at a single nucleon accounts for
the following factor 1/2. 
The factorial $3!$ occurs because 
the outgoing nucleons are indistinguishable. 
The first sum (including $1/8$) averages over the initial
spin-projection quantum-numbers and runs over all outgoing
spin and isospin states while 
the second sum contains the $4\times 36$ diagrams mentioned above
which are numerically evaluated including all interference terms. 
The integration in Eq.~(\ref{sigma}) is carried out over the Lorentz
invariant phase space spanned by the momenta 
of the outgoing particles,
$  d{\bf L}_{inv} =  \prod\nolimits_f d^3p_f/(2p_f^0) \;
           \delta^4(\sum p_f - \sum p_i)
$.

\section{On-shell problem}

In some regions of the phase space the intermediate
particle $X$ can become on-shell and the propagator gets
singular. In the following we discuss first the situation
for the pion propagator $\pi_2$ in Fig.~\ref{graph1}
which in a large part of the phase 
space  becomes  on shell.  
The standard procedure 
\cite{chung1,danielewicz,batko,peierls} 
to avoid the singularity is to introduce into the
propagator an imaginary part $im\Gamma$. 
That means that one writes the 
matrix element of Fig.~\ref{graph1} as
\beq
  {\cal{T}} =  \tilde{T_{1}} \,\frac{1}
               {p^2-m^2+im\Gamma}\, \tilde{T_2}, 
               \label{propaux}
\eeq
where the quantities $\tilde{T_1}$ and $\tilde{T_2}$
 describe the two subprocesses
1+2 $\to$ 4+5+$\pi$ and  $\pi$+3 $\to$ 6+$\phi$,
which are connected by the propagator of the
pion with mass $m=m_\pi$ and four-momentum $p^\mu$.
The physical reason for the occurrence of the imaginary part 
is that the intermediate particle does not propagate
in vacuum but in a dense medium instead. Then it can move only
a finite distance of the mean free 
path $\lambda$ during its life time $\tau$ which is related to
the imaginary part in the denominator via $\Gamma = 1/\tau$ 
(see refs.~\cite{danielewicz,batko}). 
The same method is applied if unstable particle are treated
in the entrance channels \cite{chung1,peierls}.

However, the on-shell propagation is in fact a process which is
already treated in standard transport models 
as two consecutive two-step processes, where
the particle $X$ is treated as a real on-shell particle. 
Now we try to separate the genuine three-body 
processes out of the contributions given by the
whole propagator in Eq.~(\ref{propaux}). 

If the particles were in vacuum 
the amplitude ${\cal{T}}$  would also be given 
by Eq.~(\ref{propaux}),  but the quantity
$m\Gamma$ would be replaced by $ \varepsilon$
with Feynman's prescription $ \varepsilon \to +0$.
The square of this
matrix element diverges since it is proportional
to the square of the  function $\delta(p^2-m^2)$ in the 
limit $\varepsilon\to 0$. Since the origin 
(see e.g. \cite{itzykson}) of the
$\delta$ function is the integration over the
time the pion moves,  one can 
replace it  with $\tau \delta(p^2-m^2) /(2 \pi m)$,
where the time $\tau$ is connected with $\varepsilon$
via $\tau = m/\varepsilon$.
Thus, the transition probability is proportional to the
proper time $\tau$ in the rest system of the 
intermediate particle. 
This treatment is analogous to the standard 
derivation \cite{itzykson} of the cross section 
(\ref{sigma}) which becomes proportional
to the space-time volume. 

Due to the terms coming from the exchange of the initial and
final momenta the matrix $\cal{T}$ 
becomes on shell in different regions  of the phase space
spanned by the four final-state  momenta. 
In general there are 3 $\times$  3 subspaces defined 
by $(p_a-p_b)^2=m^2$, where $p_a$ and $p_b$ are the total momenta
of the three possible initial pairs $a$ =[(12),(23),(31)] and three
final pairs $b$ = [(4,5),(56),(64)], respectively. 
 
In the near neighbourhood of each subspace we divide the 
 ${\cal{T}}$ matrix in a smooth part and a singular one,

\beq
      {\cal{T}} = \bar {\cal{T}} +  T_1 \, \frac{1} 
                   {p^2-m^2+i\varepsilon} \, T_2 \;.
        \label{divide}
\eeq
Contrary to Eq.~(\ref{propaux}) the amplitudes $T_1$ and 
$T_2$ describe now subprocesses with free (on-shell) particles. 
In calculating the phase space integral 
we treat the square of the second term as
\beq
 \label{sigma12}
\sigma' &=& \sigma_{two-step} \, + \, \sigma'_{three}  \;,\\
 \label{sigma2}
\sigma_{two-step} &=& C {\displaystyle \int} d{\bf L}_{inv} \;
             \frac{\pi\tau}{m} \delta(p^2-m^2)\; |\hat{T}|^2 \;,\\
\sigma'_{three}  &=& C \,  \lim\limits_{\varepsilon\to 0}
            {\displaystyle \int} d{\bf L}_{inv} \; |\hat{T}|^2
 \big[\frac{(p^2-m^2)^2}
                 {[(p^2-m^2)^2 +\varepsilon^2]^2}
                 -\frac{\pi}{2\varepsilon}\delta(p^2-m^2)\big],
         \label{sigma3}
\eeq
where $\hat{T}$ = $T_1T_2$.
The quantity $C$ comprises the factors standing in front
of the integral in Eq.~(\ref{sigma}).
These expressions arise by splitting the propagator  
in Eq.~(\ref{divide}) into the $\delta$ function and 
its principal value, while the last term $\sigma'_{three}$ is 
constructed such that
it is finite as we will see below.

We identify the first part (\ref{sigma2})
with the two-step process as it is proportional to the
time of flight of the intermediate particle which can
only be fixed by additional information on the collision geometry.
This part can be transformed into  an expression describing
a sequence of two  two-nucleon collisions.  
Introducing the identity $\int d^4p \, \delta^4(p_1+p_2-p_4-p)=1$
and using the definition of the 
two-particle cross sections we obtain
\beq
\sigma_{two-step}=   \int d{\bf p}_\pi \;
\frac{d\sigma}{d{\bf p}_\pi}({\bf p}_1,{\bf p}_2) \;
\sigma_\phi({\bf p}_\pi,{\bf p}_3)\; (n_0 \hat{v}\tau) \;.
\label{twostep}
\eeq

Now the cross section has been factorized in the cross sections 
 $d\sigma/d{\bf p}_\pi$ for pion production 
and $sigma_\phi$ for $\phi$ production via pion absorption 
by the nucleon 3. The sum over the final
subspaces $b$ has cancelled the factorials in the definitions
of the cross sections.
In principle one should sum Eq.~(\ref{twostep}) over the 
three subspaces $a$, however,  we have used only the first one
in accordance to Eq.~(\ref{sigma}). 
The last factor in the round bracket contains 
the relative velocity $\hat{v} =(p_\pi p_3) v_{rel}$, 
$ v_{rel}= \sqrt{(1- (mm_N/(p_\pi p_3))^2}$ 
between the intermediate particle
and its partner 3. It originates from the flux factor 
in the definition of the production cross section 
and is proportional to the length $L=\hat{v} \tau$ the pion travels. 

A similar expression like Eq.~(\ref{twostep}) arises if the 
intermediate nucleon $X$ moves on shell.
Formally, the integral over the pion momentum in Eq.~(\ref{twostep})
is replaced by an integral over the scattering angle of the 
elastic nucleon-nucleon scattering. To derive such an equation 
the on-shell part of the amplitude $\hat{T}$, see Eq.~(\ref{sigma2}),
is transformed into a sum over the spin quantum numbers of the 
intermediate nucleon using the properties of the Dirac operator,
\beq
                \hat{T} = {T_{1}} 
               (\gamma^\mu p_{\mu} +m_N) {T_2}
               =\sum_n T_{1n}T_{2n} \;.
\eeq
To arrive at the standard formula the following approximation
is needed: $\sum |\sum_n T_{1n}T_{2n}|^2 \approx
1/2 \sum|T_1|^2 \sum|T_2|^2$ which is fulfilled
if spin-flip processes are not important.

If we apply Eq.~(\ref{twostep}) to a p+A collision of a nucleon 
on a nucleus of 
mass number $A$ the cross section would be proportional to 
$A^{4/3}$ in the subthreshold region as the length $L$ is proportional
to $A^{1/3}$.
However, if the reaction 
proceeds in a dense medium the length reduces to the mean free 
path, and the factor $n_0 \hat{v} \tau = 1 / \sigma_{tot}$
gives the inverse total cross section. 
This brings us back to the prescription of Eq.~(\ref{propaux})
and the two-step formula employed, e.g., in \cite{sibirtsev}.
In this case the cross section of a p+A reaction is proportional
to $A^{2/3}$ taking into account the fact that the incoming particle
suffers rescattering processes too.

After separating the two-step process from the 
total cross section the remaining parts
form the genuine three-body cross section $\sigma_{three}$.
It contains the smooth terms of the $\cal{T}$ matrix  in Eq.~(\ref{divide})
and the limit of the divergent contributions of Eq.~(\ref{sigma3}).
To see that the  expression in the square brackets  in Eq.~(\ref{sigma3})
is finite we analytically integrate over $p^2$ around the singularity 
within the interval between $p^2=m^2\pm mD$. 
Assuming that $\hat{T}$ is a
smooth function of $p^\mu$ we obtain $-2/mD$  for       
a small but finite value of $D$.
Dividing the integration into parts near and far the singularity
we write the three-body cross section as
\beq
\nonumber
\sigma_{three} &\,=\,& C {\displaystyle \int} d{\bf L}_{inv}
         \bigg[ \, \Theta(|p^2-m^2|- mD) \, 
          | {\cal{T}}|^2 \\
      &&  + \delta(p^2-m^2)\; \bigg( 2mD| \bar{{\cal{T}}}|^2 +
         2\pi\, {\cal{I}}m(\bar{{\cal{T}}}\hat{T}^*)-
          \frac{2}{mD} |\hat{T}|^2\bigg)\bigg] \;,
\label{three}
\eeq
where $\Theta$ denotes the step function. 
Terms of higher than first order in $mD$ have been neglected.

\section{In-medium effects}

In Eq.~(\ref{sigma3}) we have assumed that  the intermediate
particles move freely. Inside a medium we can employ 
a finite  imaginary part as already inserted in Eq.~(\ref{propaux}).
Recently great effort has been 
made to construct the in-medium propagators and
corresponding spectral functions calculating the self-energies in 
one-loop order and higher approximations to couple the
propagating particles to the constituents of the surrounding
medium. However, for simplicity reasons we treat 
these effects  phenomenologically as collision broadening \cite{bugg}
introducing in the propagator in Eq.~(\ref{propaux}) the collision width 
$\Gamma_{coll}= n_0 \hat{v} \sigma_{tot}$
\beq
   \Gamma =  \frac{1}{\tau} = 
      \Gamma_0(p^2)+ \Gamma_{coll} 
   \label{gamma}
\eeq 
We have already foreseen that the intermediate particle
is a resonance with an energy dependent decay width $\Gamma_0$. 
Although the definitions in Eq.~(\ref{sigma3}) can be used also for
a finite value of $\varepsilon$ after transforming 
back the $\delta$ functions
and their squares into the finite-epsilon representations, 
this procedure is ambiguous here since there is no
singularity anymore.
Thus, we define the genuine three-body
cross section as the difference of the total cross sections
and all possible two-step cross sections:
\beq
\sigma_{three} = C\int d{\bf L}_{inv} \bigg( |{\cal{T}}|^2
     \;-\; \sum_{c} |\hat{T}_c|^2 \delta(p_c^2-m_c^2) \,
        \frac{\pi}{\Gamma m_c}\bigg)\,.    \label{threemore}
\eeq
The sum runs over all open channels $c$ for 
two-step collisions, where resonances
are  included as long as their decay channels are open, e.g. 
intermediate $\rho$
mesons with masses $m_c$ being larger than 
the two-pion decay threshold.

Due to the surrounding medium the  $\cal{T}$ matrix 
looses its Lorentz invariance as the relative velocity 
refers to the medium's rest frame which we fix to be the laboratory 
system. 
We mention that the value of $\hat{v}$ becomes very large for pions
which create a $\phi$ meson. A pion needs a value of  $\hat{v}\sim 11$
in a collision with a nucleon resting in the laboratory system, 
while its cross section amounts to about 25 mb. 
The corresponding relative velocity for a $\rho$ meson
is considerably smaller due to its larger mass.   

In our calculations we use the effective widths defined 
in Eq.~(\ref{gamma}) for those 
particles only which serve as intermediate particles, and treat
particles which occur inside the diagrams of the two consecutive
processes like free particles. This corresponds to the
widely used standard method by which cross sections are calculated 
with free particle propagators.

\section{Numerical results}

Now we are going to discuss the cross section for a proton which 
hits another  proton at rest  embedded in a protonic surrounding
with density $n_0$=0.16 fm$^{-3}$. We vary  
the bombarding energy $E_{lab}$ 
from the three-body threshold  of 1.8 GeV 
for $\phi$ meson production up to the two-nucleon threshold
of 2.6  GeV. 
In this special case of two particles without relative motion
the nucleonic intermediate states cannot get on shell
below the two-nucleon threshold while 
pions and $\rho$ mesons already allow two-step processes at an 
bombarding energy of about 0.04 GeV above threshold.

\begin{figure}
\centerline{
\includegraphics[width=16cm,height=16.cm]{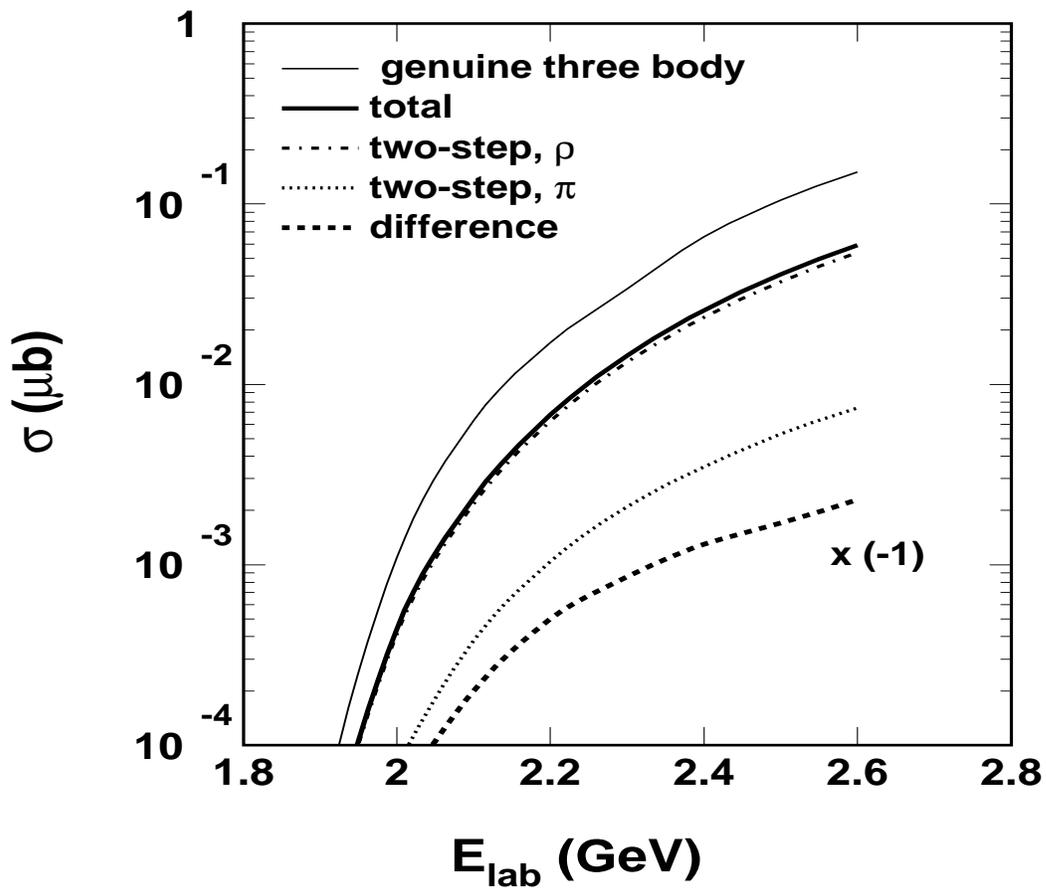}  }
\caption{
Cross section of $\phi$ meson production by a proton hitting a proton
at rest embedded within
a proton environment. The thin line gives the three-body 
cross section without
intermediate rescattering whereas the other lines show the results
in a medium with rescattering. For details see text.
\label{cross2} }
\end{figure}

First we have calculated the cross section  without the in-medium effect.
The two-step cross section (\ref{sigma2}) is proportional to the flight time
$\tau$ or the system size. 
On the other hand the genuine three-body 
cross section which is calculated according 
to Eq.~(\ref{three}) does not depend on the system size
and is presented in Fig.~\ref{cross2} by the thin full line. 

The cross sections are considerably diminished if the rescattering
effects of the intermediate particles are taken into account.
We have used a total cross section of 25 mb for both pions and $\rho$ mesons.
The two corresponding two-step processes  
(defined within the right hand site of 
Eq.~(\ref{threemore}))
are also displayed by the dashed and the 
dot-dashed lines  in Fig.~\ref{cross2}, respectively. 
These cross sections  are considerably smaller
than the original three-body cross section. The sum of 
these two quantities has to be compared to the total cross section
calculated with the complete $\cal{T}$ matrix of Eq.~(\ref{threemore})
which is presented by the
thick full line. The difference should be the contribution of the
genuine three-body cross section which, however,  
turns out to be negative (thick dashed line).
The reason is that due to the large width of
200 MeV the pion propagator in the three-body
matrix element reaches out very far in the phase space and picks up
$\cal{T}$ matrix elements which are smaller than in the on-shell region.

\begin{figure}
\centerline{
\includegraphics[width=16cm,height=16.cm]{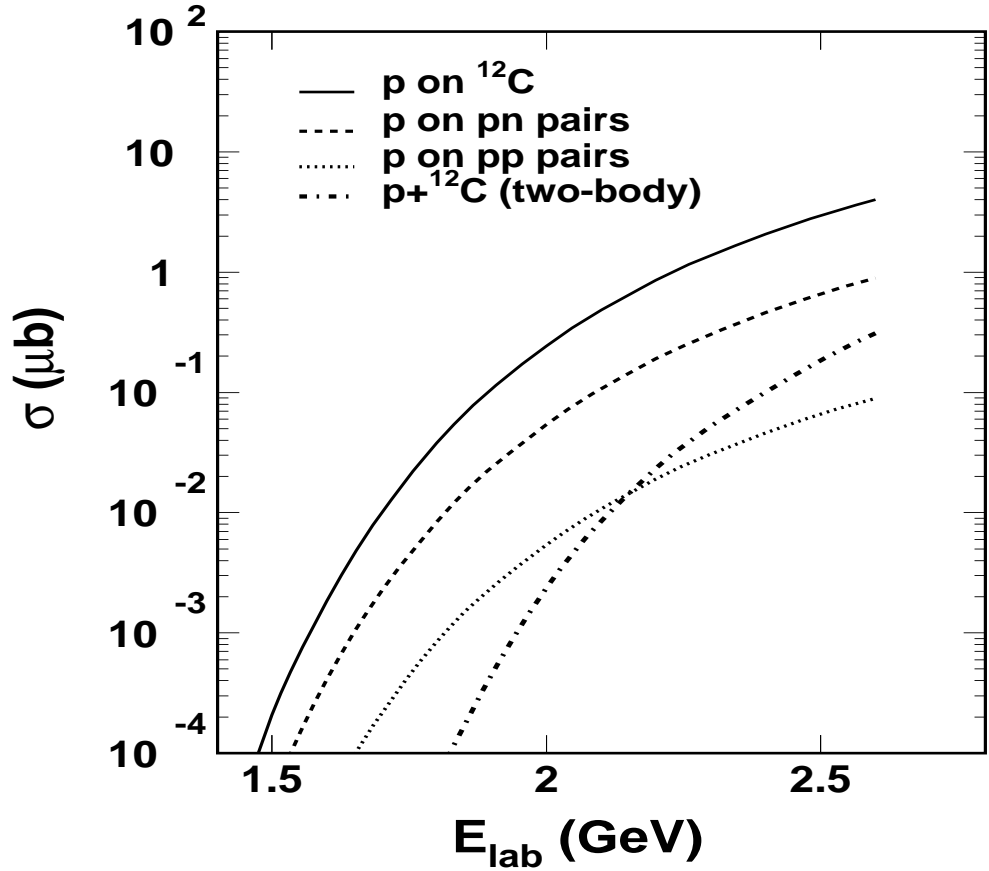}  }
\caption{
Cross section for $\phi$ meson production for a proton on carbon 
collision (solid curve). Shown are also the cross sections for 
a proton colliding with a proton surrounded by protons (dotted line) 
or neutrons (dashed line) including  Fermi motion. The dash-dotted line 
shows the effect of two-nucleon collisions. 
\label{cross3} }
\end{figure}

Finally, we calculate the cross section  for the collision of
a proton on a target nucleus consisting of $N$ neutrons and $Z$ protons.
The production is now a superposition of primary 
reactions on  pp and pn pairs. 
The production at a pn pair is strongly
enhanced compared with that at a pp  pair because the isospin coupling
allows in the latter case only $\pi^0$ exchanges which have
smaller vertex strengths than those of charged mesons. In more complete
calculations which,  e.g., include also the exchange of $\sigma$ mesons 
such a large difference may not occur. 
Furthermore, the nucleons have Fermi momentum which can be taken 
into account by using the  spectral function  
\mbox{$\delta(p^0 - m_N - {\bf p}^2/[2(A-1)m_N] + E_{sep})
\exp{(-5{\bf p}^2/2k_{Fermi}^2)}$} with a Fermi momentum of
$k_{Fermi}$ = 0.27 GeV and an average  separation energy of 
$E_{sep}$ = 20 MeV.
The effect of the Fermi motion allows the 
production already at much lower energy.
In view of the exploratory nature of our calculations
it is not worth using improved spectral functions which can be found
in ref.~\cite{atti}. Due to the effect of Fermi motion 
the production already begins at much lower energy. 
In Fig.~\ref{cross3} the
cross sections $\sigma_{p-pp}$  and $\sigma_{p-pn}$ 
for $\phi$ production at a proton  in the surrounding of protons or
respectively neutrons are presented. 
They are used to calculate the $\phi$ production
in the reaction of a proton with  $^{12}$C in accordance to
 $\sigma = [(Z(Z-1) \sigma_{p-pp} + 2ZN \sigma_{p-pn}
+N(N-1) \sigma_{p-nn}]/A$. 
We have used an effective particle number $A$ = 6 
as in ref.~\cite{sibirtsev} 
which results from the fact that many target nucleons
are screened due to scattering processes of the incoming proton.
For comparison we have also included into Fig.~\ref{cross3}
the $\phi$ production cross-section
(dash-dotted line) calculated as superposition of primary 
two-nucleon collisions, 
$\sigma = Z \sigma_{p-p} +
N \sigma_{p-n}$,   which is smaller than the three-body 
production, in contrast to the results of ref.~\cite{sibirtsev}.

Finally we should keep in mind that we have constrained our 
Lagrangian of  the nucleon-nucleon-meson interaction. For instance
neglecting the intermediate $\Delta$ particles 
causes an underestimation of the pion production. 
Therefore,  the actual $\phi$ production
is expected to be larger than presented in Figs.~\ref{cross2}
and ~\ref{cross3}.

\section{Conclusion} 

It has been our aim to study the features and consequences of 
elementary three-body collisions.
This is an important issue when calculating particle production
near threshold for nucleon-nucleus and nucleus-nucleus collisions
within the frame work of transport models. 
Here, we have studied the $\phi$ production below
the free nucleon-nucleon threshold.
It turned out that genuine three-body processes are not important
as most of the intensity can be evaluated via consecutive two-nucleon
reactions. In the case considered here the two-step processes 
slightly overestimate the cross section. 
However, it is necessary that all intermediate particles
which can become on shell are treated properly 
within the transport codes as they contribute essentially to particle
production.

\newpage


\newpage


\end{document}